\documentclass[english,prd,letter,twocolumn,showpacs]{revtex4}
\usepackage{graphicx}
\begin{document}

\newcommand{\ahubble}{\,\frac{\dot{a}}{a}\,}
\newcommand{\overa}{\,\frac{1}{a}\,}
\newcommand{\uvec}{\mathbf{u}}
\newcommand{\kvec}{\mathbf{k}}
\newcommand{\bfnabla}{\mathbf{\nabla}}
\newcommand{\bfbfield}{\mathbf{B}}
\newcommand{\bfcdot}{\mathbf{\,\cdot\,}}
\newcommand{\cs}{c_{\mbox{\scriptsize{S}}}}
\newcommand{\ca}{c_{\mbox{\scriptsize{A}}}}
\newcommand{\fpi}{4\pi}
\newcommand{\valfven}{\mathbf{v}_{\mbox{\scriptsize{A}}}}
\newcommand{\upla}{\mathbf{u}_{\mbox{\scriptsize{P}}}}
\newcommand{\rhop}{\rho_{\mbox{\scriptsize{P}}}}
\newcommand{\Jplas}{\mathbf{J}}
\newcommand{\jfield}{\mathbf{J}}
\newcommand{\mfi}{\mathbf{B}}
\newcommand{\bfield}{\mathbf{B}}
\newcommand{\efi}{\mathbf{E}}
\newcommand{\ung}{\mathbf{u}_{\mbox{\scriptsize{NG}}}}
\newcommand{\rhong}{\rho_{\mbox{\scriptsize{NG}}}}
\newcommand{\udm}{\mathbf{u}_{\mbox{\scriptsize{DM}}}}
\newcommand{\rhodm}{\rho_{\mbox{\scriptsize{DM}}}}
\newcommand{\mnras}{MNRAS}

\title{A Cautionary Note on Cosmological Magnetic Fields}

\author{Lu\'{\a i}s F. A. Teodoro, Declan A. Diver and Martin A. Hendry}

\affiliation{
  Department of Physics and Astronomy\\
  Kelvin Building\\
  University of Glasgow \\
  Glasgow, G12 8QQ, Scotland, UK
}%
\email[Enquiries to ]{luis@astro.gla.ac.uk}
\date{\today}%

\begin{abstract}
This note is concerned with potentially misleading concepts in the
treatment of cosmological magnetic fields by magnetohydrodynamical
(MHD) modelling. It is not a criticism of MHD itself but rather a
cautionary comment on the validity of its use in cosmology. Now
that cosmological data are greatly improved compared with a few
decades ago, and even better data are imminent, it makes sense to
revisit original modelling assumptions and examine critically
their shortcomings in respect of modern science. Specifically this
article argues that ideal MHD is a poor approximation around
recombination, since it inherently restricts evolutionary
timescales, and is often misapplied in the existing literature.
\end{abstract}


\maketitle
\section{Introduction}
The role of the magnetic field in the evolution of the Universe
has been a challenging problem for many decades, retaining a
frustrating degree of speculation despite remarkably inventive and
ingenious mathematical modelling. Definitive progress to date has
been hampered by the ambivalence of observational data: as yet,
contradictory hypotheses continue to remain possibilities for as
long as the definitive observation remains elusive.

However, the quantity and precision of appropriate cosmological
data are soon to be revolutionised with the imminent construction
of dedicated instruments such as SKA, and the era of precision
diagnosis of cosmological magnetic fields is not far away. It is
surely appropriate then to re-examine the theoretical basis upon
which the modelling of cosmological magnetic fields is based, in
preparation for these new data.

A favourite mathematical framework for modelling the evolution of
magnetism near the decoupling era (and sometimes for the early
universe) is magnetohydrodynamics (MHD), which blends the continuum
properties of the matter in the Universe with a restricted set of
electromagnetic concepts, yielding a hybrid magneto-fluid context in
which to study the influence of matter on magnetic fields, and
vice-versa. Such MHD models are often imbued with perfect electrical
conductivity (that is, zero electrical resistivity), and are almost
invariably single-fluid, since this combination allows significant
simplification of mathematical modelling, without apparently
sacrificing too much of the underlying physics. It has to be
mentioned that many of the cosmological MHD models used in this
context have additional properties not encountered in standard
plasma modelling; these primarily arise from the incorporation of
Universal expansion, and sometimes from accommodating radiation
pressure as an extra force term.

In this short article we examine critically the physical
underpinning of the magnetofluid Universe as a suitable model for
cosmological magnetogenesis and evolution, and find such models to
be flawed in basic concept. We conclude that MHD, and its
variants, are quite unsuited to modelling the critical phenomena;
instead we believe that consistency demands that we use a
relativistic kinetic theory instead
[see \cite{Berstein:1988kteu.book.....B} for a gas kinetics (but not plasma) treatment
in a Friedman-Lema\^{\i}tre-Robertson-Walker (FLRW) space-time].
The next sections explain our
reasoning, by highlighting the inadequacies of the magnetofluid
approach, ranging over issues such as the calculation of
conductivity, the meaning of the current density, the
interpretation of mass density and the allied velocity field, and
finally comments on the juxtaposition of relativity and
electromagnetism with an MHD plasma description.

\section{The relevance of MHD to Cosmology}

Magnetohydrodynamics is a fluid model, in which aspects of
electromagnetism are incorporated into standard hydrodynamics to
give an electrically conducting continuum that can generate a
self-field, and respond to an applied one. Single fluid MHD cannot
support charge separation, and perfectly conducting (or ideal)
MHD has zero electric field in the rest frame of the plasma.
\par

MHD is the favoured mathematical model for the description of the
cosmic plasma behaviour but caution is required in the
interpretation of the results. MHD is an excellent plasma model in
many disparate contexts, but care has to be taken to ensure that
conclusions are not drawn that cannot be sustained by the
restricted physics implicit in MHD, particularly where the context
requires elements of relativity.

In cosmological MHD there may be at best a conflict in
terminology, and at worst, flawed physics. In order to be clear
about the problems, we elaborate below on general issues in which
the physical interpretation is perhaps at odds with the
mathematical framework.

One fundamental aspect in cosmological descriptions involving
plasma is as follows: in the literature it is assumed that all the
universe is ionised, meaning the plasma density and baryon density
are identical. This may be reasonable before recombination
(assuming that dark matter can be ignored) but not after! Whilst
this might appear at first a trivial statement, the implications
can be significant if care is not taken to identify how the mass
densities of constituent fluids contribute to the relevant
physical forces.

Below we present a list of issues in which the use of MHD may lead
to inconsistencies in the interpretation of the results. These are
not in any particular order, and we don't claim that it is a
exhaustive list. However, each topic highlights a specific
conceptual shortcoming associated with the use of MHD in
cosmological descriptions currently in the literature.

\subsection{The Perfectly Conducting Assumption} One of the great
strengths of single-fluid MHD is the simplicity in the field
evolution, particularly given the ideal case in which the plasma
has zero resistivity. Given that the single-fluid model
immediately rules out charge separation, the further imposition of
a perfectly conducting fluid allows a very simple Ohm's law to be
placed at the heart of the physics:
\begin{equation}
\mathbf{E} + \mathbf{u}\times\mathbf{B} = 0 \label{exb}
\end{equation}
where $\mathbf{E}$, $\mathbf{u}$ and $\mathbf{B}$ are respectively
the plasma electric field, bulk fluid velocity and magnetic flux
density (loosely, magnetic field). One consequence is immediate:
the electric field can only be produced as the result of a frame
change, and is therefore entirely prescribed by the velocity and
magnetic fields, having no independent evolution. (There are no
induced fields, for example.) This is because any applied electric
field invokes a plasma dynamical response in terms of the particle
current density $\mathbf{J}$ consistent with the relation
\begin{equation}
\mathbf{J}=\mathbf{\sigma}\mathbf{E} \label{condc}
\end{equation}
where $\mathbf{\sigma}$ is the conductivity tensor; not a scalar,
since the magnetic field changes the transport properties of
mobile species for cross-field motion compared to the isotropic
case. Clearly then unlimited particle currents can be sustained in
the perfect conductivity case by vanishingly small electric
fields; since such currents are not seen, then the Universal
electric field must be zero except for that arising from a frame
translation of the magnetic field. Whilst this is perfectly
acceptable within the classical concept of MHD plasmas, it pays to
consider carefully the implications of this in a cosmological
context.

In the early Universe (at least pre-decoupling), the plasma
(charged baryons and electrons) co-exists with a radiation field
that interacts with it via the Thompson scattering of photons by
the free electrons (to identify but one process). Viewed
classically (that is, non quantum-mechanically) in terms of the
propagation of the electric field disturbances, the plasma is a
dielectric, since interaction between the two continua produces a
different response than if the radiation simply propagated in a
classical vacuum: the plasma has a refractive index different from
unity, the classical vacuum value. A non-vacuum refractive index
implies a finite conductivity. This can be seen from cold plasmas,
for example, in which the dielectric tensor $K$ and conductivity
tensor $\sigma$ are intimately related \cite{Boyd:1969pldy.book.....B,Stix:1992}
\begin{equation}
K=I+\frac{i}{\epsilon_0\omega}\mathbf{\sigma}
\end{equation}
where $I$ is the unit tensor, $\omega$ is the frequency of the
electromagnetic radiation, and $\epsilon_0$ is the vacuum
permittivity. Note that the kinetic plasma treatment is similar.
Hence if the plasma refractive index is not unity (the classical
vacuum value), then the conductivity of the plasma must be finite,
and anisotropic: particle transport parallel to the magnetic field
is different from that perpendicular to it.

Of course, in the very early universe where current carriers are
effectively massless then the conductivity could be very
high \cite{Ahonen:1996PhLB..382...40A,Baym:PhysRevD.56.5254}; however,
the applicability of MHD in such energetic
conditions is very much open to question.

The transition from perfect to finite conductivity is subtle, but
critical:  finite conductivity brings evolutionary and topological
constraints that are simply missing from perfect conductivity;
moreover a resistive plasma has additional interdependence of the
fields compared with an ideal one.

Note that in the recombination era, the conductivity must evolve
as the charged particle density changes, making any assumption of
perpetual perfect conductivity even less credible (see the
discussion later in this section).

Hence a more appropriate general form of Ohm's Law is the simplest
resistive one,
\begin{equation}
\mathbf{E} + \mathbf{u}\times\mathbf{B} = \eta \mathbf{J}
\label{Ohm}
\end{equation}
where $\eta=\sigma^{-1}$ is the plasma resistivity, also a tensor
in a magnetised plasma.

It is a cosmological convention to calculate electrical
conductivity in terms of the effect of photon scattering on
electron transport, using a simple scalar expression for the
conductivity based on a single interaction time $\tau$
\cite{Grasso:2001}:
\begin{equation} \sigma = \frac{n_e e^2
\tau}{m_e}\label{BGK}
\end{equation}
in which $\tau=1/(n_\gamma \sigma_T )$, with $n_\gamma$ is the
photon number density and $\sigma_T$ is the Thomson  scattering
cross-section, and $n_e$ is the \textit{free}-electron number
density. This yields a post-recombination conductivity that is
approximately $10^{-3}$ times that for the pre-recombination case
\cite{caprini-2005-0502}. Taking the plasma perspective, the ratio
of electrical conductivities before and after recombination is based
on the Spitzer electrical conductivity for a fully ionized plasma at
a temperature of 10~eV \cite{DADbook:2001}, and the scattering of
0.1~eV electrons by neutral hydrogen \cite
{HipplerPfauSchmidtSchoenbach2002/026,Buckman:1985JChemPhys82.4999}.
This ratio is
\begin{equation}
\sigma_{\mbox{\scriptsize{pre}}}/\sigma_{\mbox{\scriptsize{post}}}\approx 0.016/ x_e
\end{equation}
where $x_e$ is the ionization fraction. This yields a similar drop
in conductivity as in the Thomson scattering case if $x_e \approx
10^{-5}$\cite{Padmanabhan:1993}. Since the Spitzer conductivity is
similar in value to that calculated by considering only the Thomson
scattering of photons by electrons then this plasma calculation
shows that the pure plasma treatment is equally valid, though the
nature of the BGK approximation in calculating the resistivity
Eq.~(\ref{BGK}) might not be wholly valid in either context, since
it is unlikely that a single relaxation time can account for all
scattering processes. Either way, the resistivity is finite on
either side of recombination, and significantly lower afterwards,
underlining the dangers in a perfectly conducting description.

\subsection{The Total Current}
In the present article we use the $1+3$ covariant formalism
introduced by
\cite{Ellis:1973CLP.....6....1E,Ellis:1989PhRvD..40.1804E}, in
which a general class of homogeneous space-times are considered.
The usual $\nabla$ and $t$ operators are generalised
appropriately.

One of the crucial approximations in MHD is that the current
density is given exclusively in terms of the magnetic field
curvature:
\begin{equation}
\nabla \times \mathbf{B}=\mu_0\mathbf{J} \label{curlbmhd}
\end{equation}
Note that $\mathbf{J}$ is not related directly to the bulk flow
velocity $\mathbf{u}$, since the latter is a mass-weighted
averaged quantity that is independent of charge. Hence in MHD, the
current density is a dependent variable, and can be eliminated
everywhere in favour of the magnetic field.

However, full electromagnetism demands that
\begin{equation}
\nabla \times
\mathbf{B}=\mu_0\mathbf{J}+\frac{1}{c^2}\frac{\partial
\mathbf{E}}{\partial t} \label{curlbem}
\end{equation}
where $c$ is the speed of light, and $t$ is the time. The
additional term on the right is the displacement current, omitted
from MHD because the rapid time evolution of electromagnetic
effects is not incorporated in such a fluid description: there are
other comparable time-dependent terms in the accompanying plasma
equations that have also been
omitted\cite{Boyd:1969pldy.book.....B} and so, for mathematical
consistency, the displacement current must be dropped.

The true significance of adopting Eq.~(\ref{curlbmhd}) instead of
Eq.~(\ref{curlbem}) lies in the timescale for the evolution of
the magnetic field. With Eq.~(\ref{curlbmhd}) we have
\begin{eqnarray}
\frac{\partial {\mathbf{B}}}{\partial
t}&=&-\nabla\times\mathbf{E} - \frac{2}{3}\Theta\mathbf{B} \\
&=& \nabla\times(\mathbf{u}\times\mathbf{B}-\eta\mathbf{J}) - \frac{2}{3}\Theta\mathbf{B}
\end{eqnarray}
where we have used Eq.~(\ref{Ohm}) for the most
general case. $\Theta$ denotes
$3\dot{a}/a$, where $a$ is the expansion factor.
Eliminating the current density using Eq.~(\ref{curlbmhd}) yields
\begin{equation}
\frac{\partial \mathbf{B}}{\partial
t}=\nabla\times(\mathbf{u}\times\mathbf{B})+\frac{\eta}{\mu_0}\nabla^2\mathbf{B} - \frac{2}{3}\Theta\mathbf{B}
\end{equation}
assuming the simplest case of constant resistivity. This parabolic
equation shows that the magnetic field evolves on the resistive
diffusion time-scale (notwithstanding the feedback term from the
dynamo contribution, the first term on the right-hand side);
cosmological expansion will allow an additional cosmological
time-scale.

However, if $\mathbf{J}$ is eliminated using Eq.~(\ref{curlbem})
instead, the governing equation for the magnetic field is
higher-order, and hyperbolic:
\begin{equation}
\frac{\eta}{\mu_0c^2}\frac{\partial^2\mathbf{B}}{\partial
t^2}+\frac{\partial\mathbf{B}}{\partial t}=\nabla \times
(\mathbf{u}\times\mathbf{B})+\frac{\eta}{\mu_0}\nabla^2\mathbf{B} - \frac{2}{3}\Theta\mathbf{B}
\end{equation}
This introduces new time-scales for field evolution, including
wave propagation, damping and growth, to be balanced against the
dynamo term and any cosmological input.

Granted, magnetic and electric fields that so arise may be rapidly
varying: this means that the overall mathematical framework must be
able to accommodate fast time-scale evolution, a point to which we
will return in a different context later in this article.

Note that in some very early work
(e.g.\,\cite{Wasserman:1978ApJ.224.337W}) the magnetic structure
was modelled without any intrinsic time evolution, and only varied
on cosmological time scales; other more recent work omits the
spatial structure of the magnetic field
(e.g.\,\cite{caprini-2005-0502}). Neither of these approaches is
correct: the true picture must have the correct balance between
spatial and temporal derivatives in order to reflect the correct
physics.

\subsection{Density considerations}
One serious concern with unified fluid descriptions of the
universe around recombination is the concept of the mass density.
Care must be taken to distinguish between the density of charged
and non-charged particles; the latter are also divided between
baryonic matter and dark matter. Only the density of plasma (that
is, the electrically conducting and magnetised fluid) can appear
in those plasma equations that are concerned with exclusively
plasma effects. For example, neutral matter density cannot
contribute to current densities. There are several examples in the
literature
\cite{Subramanian:1998PhRvL.81.3575S,Subramanian:1998PhRvD.58h3502S,Jedamzik:1998PhRvD.57.3264J}
where only baryonic matter density enters in the model equations
with no distinction between neutrals and current-carrying species.
Where baryonic matter is predominantly charged, so that the
baryonic and plasma matter densities are approximately the same,
this is fine, but where neutral matter (e.g. hydrogen atoms and
molecules) co-exists with plasma, this clearly cannot be correct,
and care must be taken to distinguish between species. This latter
case leads for example to an Alfv{\'e}n speed, $c_A=B/\sqrt{\mu_0
\rho_0}$, that incorrectly uses the total mass density, instead of
just the plasma mass density; the magnetic field cannot be
influenced directly by the motion of neutral matter and
vice-versa. This becomes particularly important when the
Alfv{\'e}n speed is used to estimate magnetic field strengths in
the post recombination era (for example,
\cite{Subramanian:1998PhRvD.58h3502S}).

Momentum exchange between the fluids is obviously
physically correct, and is a primary method of causing indirect interaction between the magnetic field
and neutral particles,  but is predicated on making the distinction
between the different fluid types in the model
\cite{Marklund:2003CQGra..20.1823M, Diver:2006NJPh....8..265D}.

In particular, the pressures of neutral matter
and plasma evolve under different physical conditions. Hence, incorporating the magnetic field into
the Jeans length requires a careful definition of the concept of magnetic and kinetic pressure. The latter
is influenced by both neutral matter and plasma; the former is only directly supplied by the plasma. To
get a true holistic picture, the neutral gas and plasma components must be identified as separate species
from the outset, with the latter obeying additional force terms, but with each potentially interacting
through gravitation, or explicit coupling terms.

\subsection{The Global Velocity Field}
 This problem is related to the  density issue above. Since the plasma and the neutral components
 respond to
 different dynamical equations (notwithstanding coupling) then clearly the velocity fields are
 different. In principle the neutrals need not have the same velocity as the plasma since they
 do not react directly to magnetic forces. Moreover, the bulk
 motion of neutrals makes no direct contribution to the
 evolution of magnetic field. Of course, should the neutral gas
 and the plasma be coupled in some way, then the evolution of
 one will affect the other, but this is somewhat different from
 assuming that the plasma and neutral components are somehow locked
 together.

 \indent In the light of comments about the densities and velocities it is appropriate to mention how
 plasma and non-plasma can be coupled to give an overall collective response. Plasmas and
 neutrals can be  coupled through momentum transfer \cite{Marklund:2003CQGra..20.1823M, Diver:2006NJPh....8..265D},
in which each fluid exerts a drag on the other
 by virtue of relative motion. Ionisation and recombination can also be
 considered as coupling mechanisms, in that species are
 converted from neutral to plasma and vice-versa. However single fluid MHD is not a good model for such processes,
 since charge separation is a basic pre-requisite but is impossible in single-fluid MHD plasma.
 Also, dark matter, neutrals and plasma move under the
 common self-gravitating potential presented by their respective mass
 densities, leading to dynamical equations for each species that are
 separate but coupled, emphasising that care is required in identifying what is meant by velocity at a given
 point in space since we have to distinguish between different species.

\subsection{Relativity, MHD and Photons}

In many cosmological contexts relativity is a key element of
 the physical model, including the plasma. However, there is a
 problem here if the plasma is described by a standard fluid MHD
 model. Given that MHD is exclusively concerned with low
 frequency, long wavelength phenomena that are not
 electromagnetic in nature, it is far from obvious how an MHD
 plasma description can be incorporated into a Lorentz invariant model of the whole
 ensemble: MHD cannot be Lorentz invariant, since the displacement
 current has been omitted, and so  the current density in MHD is not a Lorentz invariant.

Furthermore, photons and free electrons are formally irreconcilable with 
an MHD prescription since MHD is not fully electromagnetic and is single 
fluid: electron and ions are combined in an averaged description. Two-fluid 
MHD is a possible approach but requires, amongst other things, a more sophisticated Ohm's
law than appears in the literature. (Recall that single fluid MHD cannot sustain 
charge separation.) Hence, photon scattering using Thomson cross-sections cannot
be rigorously quantified in a single MHD context since the electron number density 
cannot formally be deduced independently, and the physics of the scattering
processes involving photons and free electrons is not consistent with 
the exclusively long time-scale processes that are valid in MHD. 
Any formal attempt to combine both approaches in a unified treatment 
of the damping of MHD waves (for example in \cite{Jedamzik:1998PhRvD.57.3264J})
is compromised by this mismatch.

On a more general point, full electromagnetic boundary conditions are not 
appropriate for MHD since the lack of displacement current means the MHD is
pre-Maxwell; this should also be taken into account in any scattering description.

\section{Conclusions}
In this article we have illustrated some shortcomings of the
magnetofluid modelling of the universe when used to determine
magnetic field contributions to cosmic development. MHD
necessarily can only describe slow timescale effects, simply
because that is its mathematical and physical basis. It is no
surprise that MHD descriptions yield slow evolutionary behaviour;
the model is incapable of delivering anything else.

Assessing the
influence of the magnetic field in post-decoupling cosmic
structure based on such models is therefore at best approximate,
and at worst very misleading. It is now timely to revisit the
fundamental basis of magnetogenesis and evolution in order to
extract the best possible interpretation of the observations.

This article is designed to provoke discussion: we cannot fully exploit
the true information content of cosmological data if the community
persists in attempting to model physical phenomena in an
inappropriate mathematical framework. Some of these contentious 
assumptions are highlighted here. We do not offer detailed
alternatives in this article, since the construction of new
modelling environments is a challenge for us all.

LFAT acknowledges the financial support of the Leverhulme Trust.
DAD and MAH are grateful to PPARC for research funding.
%

\begin{thebibliography}{10}

\bibitem{Berstein:1988kteu.book.....B}
J.~{Bernstein}.
\newblock {\em {Kinetic theory in the expanding universe}}.
\newblock Cambridge and New York, Cambridge University Press, 1988, 157 p.,
  1988.

\bibitem{Boyd:1969pldy.book.....B}
T.~J.~M. {Boyd} and J.~J. {Sanderson}.
\newblock {\em {Plasma dynamics}}.
\newblock Plasma dynamics, by T.J.M.~Boyd and J.J.~Sanderson.~London, Nelson,
  1969.~Series : Applications of mathematics series ISBN: 177616113, 1969.

\bibitem{Stix:1992}
T.H. Stix.
\newblock {\em Waves in Plasmas}.
\newblock American Institute of Physics, New York, U.S.A., rev. edition, 1992.

\bibitem{Ahonen:1996PhLB..382...40A}
J.~{Ahonen} and K.~{Enqvist}.
\newblock {Electrical conductivity in the early universe}.
\newblock {\em Physics Letters B}, 382:40--44, 1996.

\bibitem{Baym:PhysRevD.56.5254}
Gordon Baym and Henning Heiselberg.
\newblock Electrical conductivity in the early universe.
\newblock {\em Phys. Rev. D}, 56(8):5254--5259, 1997.

\bibitem{Grasso:2001}
D~Grasso and HR~Rubinstein.
\newblock Magnetic fields in the early universe.
\newblock {\em Physics Reports-Review Section of Physics Letters}, 348(3):165
  -- 266, 2001.

\bibitem{caprini-2005-0502}
C.~Caprini and P.~G. Ferreira.
\newblock Constraints on the electrical charge asymmetry of the universe.
\newblock {\em JCAP}, 0502:006, 2005.

\bibitem{DADbook:2001}
D.A. Diver.
\newblock {\em A Plasma Formulary for Physics, Technology and Astrophysics}.
\newblock Wiley-VCH, Berlin, Germany, 1st edition, 2001.

\bibitem{HipplerPfauSchmidtSchoenbach2002/026}
Rainer {Hippler}, Sigismund {Pfau}, Martin {Schmidt}, and Karl~H. {Schoenbach}.
\newblock {\em Low Temperature Plasma Physics}.
\newblock Lecture Notes in Computational Science and Engineering. Wiley-VCH,
  Berlin, 1st edition edition, 2001.

\bibitem{Buckman:1985JChemPhys82.4999}
S.~J. {Buckman} and A.~V. {Phelps}.
\newblock {Vibrational excitation of D2 by low energy electrons}.
\newblock {\em Journal of Chemical Physics}, 82:4999--5011, 1985.

\bibitem{Padmanabhan:1993}
T.~Padmanabhan.
\newblock {\em Structure formation in the Universe}.
\newblock Cambridge University Press, Cambridge, U.K., 1st edition, 1993.

\bibitem{Ellis:1973CLP.....6....1E}
G.~F.~R. {Ellis}.
\newblock {Relativistic Cosmology}.
\newblock In E.~{Schatzman}, editor, {\em Cargese Lectures in Physics}, 1973.

\bibitem{Ellis:1989PhRvD..40.1804E}
G.~F.~R. {Ellis} and M.~{Bruni}.
\newblock {Covariant and gauge-invariant approach to cosmological density
  fluctuations}.
\newblock {\em Phys. Rev. D}, 40:1804--1818, 1989.

\bibitem{Wasserman:1978ApJ.224.337W}
I.~{Wasserman}.
\newblock {On the origins of galaxies, galactic angular momenta, and galactic
  magnetic fields}.
\newblock {\em Astrophys. J.}, 224:337--343, 1978.

\bibitem{Subramanian:1998PhRvL.81.3575S}
K.~{Subramanian} and J.~D. {Barrow}.
\newblock {Microwave Background Signals from Tangled Magnetic Fields}.
\newblock {\em Physical Review Letters}, 81:3575--3578, 1998.

\bibitem{Subramanian:1998PhRvD.58h3502S}
K.~{Subramanian} and J.~D. {Barrow}.
\newblock {Magnetohydrodynamics in the early universe and the damping of
  nonlinear Alfv{\'e}n waves}.
\newblock {\em Phys. Rev. D}, 58(8):083502, 1998.

\bibitem{Jedamzik:1998PhRvD.57.3264J}
K.~{Jedamzik}, V.~{Katalini{\'c}}, and A.~V. {Olinto}.
\newblock {Damping of cosmic magnetic fields}.
\newblock {\em Phys. Rev. D}, 57:3264--3284, 1998.

\bibitem{Marklund:2003CQGra..20.1823M}
M.~{Marklund}, P.~K.~S. {Dunsby}, G.~{Betschart}, M.~{Servin}, and C.~G.
  {Tsagas}.
\newblock {Charged multifluids in general relativity}.
\newblock {\em Classical and Quantum Gravity}, 20:1823--1834, 2003.

\bibitem{Diver:2006NJPh....8..265D}
D.~A. {Diver}, H.~E. {Potts}, and L.~F.~A. {Teodoro}.
\newblock {Gas-plasma compressional wave coupling by momentum transfer}.
\newblock {\em New Journal of Physics}, 8:265, 2006.

\end{thebibliography}

\pagestyle{empty}\end{document}